\documentclass[a4paper,11pt]{article}
\usepackage{pos}
\usepackage{physics}

\title{$B\rightarrow D^{(*)}$ decays from $N_f=2+1+1$ highly improved staggered quarks and clover $b$-quark in the Fermilab interpretation}

\author*[a,b]{Pietro Butti}
\author[c]{Carleton DeTar}    
\author[d]{Aida El-Khadra}    
\author[e]{Elvira Gamiz}      
\author[g]{Steven Gottlieb}   
\author[f]{William Jay}       
\author[g]{Hwancheol Jeong}   
\author[h]{Andreas Kronfeld}  
\author[i]{Jack Laiho}        
\author[d]{Andrew Lytle}      
\author[b]{Pablo Trujillo}    
\author[b]{Alejandro Vaquero} 

\affiliation[a]{
Quantum Theory Center at IMADA and D-IAS, Southern Denmark Univ., 5230 Odense M, Denmark}
\affiliation[b]{CAPA and Departamento de Física Teórica, Universidad de Zaragoza, 50009 Zaragoza, Spain}
\affiliation[c]{Department of Physics and Astronomy, University of Utah, Salt Lake City, Utah 84112, USA}
\affiliation[d]{Department of Physics, University of Illinois, IL 61801, USA}
\affiliation[e]{CAFPE and Departamento de Física Teórica y del Cosmos, Universidad de Granada, E-18071, Granada, Spain}
\affiliation[f]{Center for Theoretical Physics, Massachusetts Institute of Technology, Cambridge, MA 02139}
\affiliation[g]{Department of Physics, Indiana University, Bloomington, Indiana 47405, USA}
\affiliation[h]{Theory Division, Fermi National Accelerator Laboratory, Batavia, Illinois, USA}
\affiliation[i]{Department of Physics, Syracuse University, Syracuse, New York 13244, USA}

\emailAdd{pbutti@qtc.sdu.dk}

\abstract{
We present an update on the analysis of semileptonic $B \rightarrow D^{(*)}$ decays at non-zero recoil. Our computation employs $2+1+1$ FNAL-MILC ensembles with highly improved staggered quark (HISQ) action for sea and light valence quarks, while the bottom quark is treated using the clover action in the Fermilab interpretation. Simulations are performed across several lattice spacings, ranging approximately from $\sim 0.15$ fm to $\sim 0.06$ fm, and for various quark masses. We will present an overview of the analysis and show some preliminary results for the form factors.
}

\FullConference{The 41st International Symposium on Lattice Field Theory (LATTICE2024)\\
 28 July - 3 August 2024\\
Liverpool, UK\\}

\begin{document}
\maketitle

\section{Introduction}
Lattice QCD calculations of meson-decay form factors are essential for high-precision Standard Model (SM) tests. Combined with experimental data, lattice results provide independent determinations of CKM matrix elements, allowing SM precision tests. Among others, $B\rightarrow D^{(*)}\ell\nu$-meson decay allows a precise determination of the $|V_{cb}|$ element of the CKM matrix and has raised a lot of interest in the recent past.\footnote{State of the art lattice results are reported in \cite{FLAG,bonn}} FNAL-MILC was the first collaboration to perform a lattice computation of the form factors at non-zero recoil with dynamical fermions, allowing for a precise determination of $|V_{cb}|$~\cite{alex}. This update talk outlines the FNAL-MILC collaboration’s progress on the analysis of the $B\rightarrow D^{(*)}\ell\nu$ for the determination of the form factors. 
In this talk, we present preliminary results of the analysis performed on the two- and three-point correlation functions aimed to extract the form factors, following the workflow of~\cite{alex}.

\section{Overview of the computation}
The correlators are computed on the FNAL-MILC $N_f=2+1+1$ ensembles generated with a 1-loop improved Lüscher-Weisz gauge action, a HISQ action for the sea and $u,d,s$ valence quarks, while $c,b$ valence quarks are treated with a clover action in the Fermilab interpretation. The whole calculation employs 7 different ensembles spanning 4 different values of the lattice spacing ($a=0.15,0.12,0.09,0.06$ fm) and 3 different values of the light-to-strange quark mass ratio ($m_l/m_s \sim 0.2,0.1$ and the physical value). \cite{ensemble}. The workflow for extracting the kinematic quantities, the overlap coefficients, and the ratio follows Ref.~\cite{alex,milcD}  which we refer for additional detail. Suitable ratios of three-point function can be used to isolate the form factors ~\cite{alex,milcD}.
 In the case of the $B\rightarrow D$ decay, we have\footnote{For the quark transition $x\rightarrow y$, we consider vector ($V^\mu_{xy} = \bar\psi_x \gamma^\mu \psi_y$) and axial ($A^\mu = \bar\psi_x \gamma^\mu\gamma_5 \psi_y$) currents.}
    \begin{align}\label{ratioD}
        &R_{+}\equiv\frac{\left\langle D(\mathbf{0})\left|V_{c b}^4\right| B(\mathbf{0})\right\rangle\left\langle B(\mathbf{0})\left|V_{c b}^4\right| D(\mathbf{0})\right\rangle}{\left\langle D(\mathbf{0})\left|V_{c c}^4\right| D(\mathbf{0})\right\rangle\left\langle B(\mathbf{0})\left|V_{b b}^4\right| B(\mathbf{0})\right\rangle} \quad
        &Q_{+}(\boldsymbol{p})  \equiv \frac{\left\langle D (\boldsymbol{p})\left|V^4\right| B(\mathbf{0})\right\rangle}{\left\langle D (\mathbf{0})\left|V^4\right| B(\mathbf{0})\right\rangle} \notag\\
        &\boldsymbol{R}_{-}(\boldsymbol{p})  \equiv \frac{\langle {D}(\boldsymbol{p})|\boldsymbol{V}| B(\mathbf{0})\rangle}{\left\langle {D}(\boldsymbol{p})\left|V^4\right| B(\mathbf{0})\right\rangle} \quad
        &\boldsymbol{x}_f(\boldsymbol{p})  \equiv \frac{\langle D (\boldsymbol{p})|\boldsymbol{V}| D(\mathbf{0})\rangle}{\left\langle D (\boldsymbol{p})\left|V^4\right| D(\mathbf{0})\right\rangle}
    \end{align}
    while those for $B\rightarrow D^{*}$ are
    \begin{align}\label{ratioDst}
        &R_{A_1}^2\equiv\frac{\left\langle D^*\left(\boldsymbol{p}_{\perp}\right)\left|A_j\right| B(\mathbf{0})\right\rangle\left\langle B(\mathbf{0})\left|A_j\right| D^*\left(\boldsymbol{p}_{\perp}\right)\right\rangle}{\left\langle D^*(\mathbf{0})\left|V^4\right| D^*(\mathbf{0})\right\rangle\left\langle B(\mathbf{0})\left|V^4\right| B(\mathbf{0})\right\rangle} \quad
        &X_V  \equiv\frac{\left\langle D^*\left(\boldsymbol{p}_{\perp}\right)\left|V_j\right| B(\mathbf{0})\right\rangle}{\left\langle D^*\left(\boldsymbol{p}_{\perp}\right)\left|A_j\right| B(\mathbf{0})\right\rangle} \notag\\
        &X_0  \equiv\frac{\left\langle D^*\left(\boldsymbol{p}_{\|}\right)\left|A^4\right| B(\mathbf{0})\right\rangle}{\left\langle D^*\left(\boldsymbol{p}_{\perp}\right)\left|A_j\right| B(\mathbf{0})\right\rangle} \quad
        &X_1  \equiv\frac{\left\langle D^*\left(\boldsymbol{p}_{\|}\right)\left|A_j\right| B(\mathbf{0})\right\rangle}{\left\langle D^*\left(\boldsymbol{p}_{\perp}\right)\left|A_j\right| B(\mathbf{0})\right\rangle},
    \end{align}
The lattice values of the ratio can be obtained to parameterize the form factors as
    \begin{equation}\label{effD}
        h_{+}(w)\equiv\sqrt{\bar{R}_{+}} \bar{Q}_{+}(\boldsymbol{p})\left[1-\boldsymbol{R}_{-}(\boldsymbol{p}) \cdot \boldsymbol{x}_f(\boldsymbol{p})\right],\quad h_{-}(w)\equiv\sqrt{\bar{R}_{+}} \bar{Q}_{+}(\boldsymbol{p})\left[1-\frac{\bar{R}_{-}(\boldsymbol{p}) \cdot \boldsymbol{x}_f(\boldsymbol{p})}{\boldsymbol{x}_f^2(\boldsymbol{p})}\right]
    \end{equation}
    for the $B\rightarrow D$ case, and 
    \begin{align}
         h_{A_1}(w) &\equiv \frac{2 \bar{R}_{A_1}}{w+1}, \quad &h_{A_2}(w)  \equiv& \frac{2 \bar{R}_{A_1}}{w^2-1}\left(w X_1-\sqrt{w^2-1} \bar  X_0-1\right),     \notag\\
         h_{A_3}(w) &\equiv\frac{2  \bar{R}_{A_1}}{w^2-1}\left(w-X_1\right), \quad &h_V(w)\equiv& \frac{2         \bar{R}_{A_1}}{\sqrt{w^2-1}} \bar X_V  \label{effDst}
    \end{align}
    for $B\rightarrow D^*$, 
    where the \textit{recoil parameter} $w$ is defined by $w = \frac{1+\boldsymbol{x}_f(\boldsymbol{p})^2}{1-\boldsymbol{x}_f(\boldsymbol{p})^2}$. After the insertion of appropriate renormalization factors, the continuum limit and the extrapolation or interpolation to physical masses have to be taken.

\section{Lattice analysis}
\subsection{2-point function analysis}
    The \(D^*\) and \(B\)-meson two-point functions are essential to extract the overlap factors and energy states, which serve as inputs for the ratio fits. The two-point functions are constructed using interpolating operators \(\mathcal{O}_{Y_a}(\mathbf{p}, t)\) containing the quantum number of the meson of interest \(Y = \{B, D^*\}\), where \(a = \{d, 1S\}\) represents the type of smearing (point and Richardson~\cite{smearing1,smearing2}), \(t\) is the time, and \(\mathbf{p}\) is the spatial momentum.
    Inserting a complete set of states between the operators in $\expval{\mathcal{O}_{Y_a}(\mathbf{p}, t)\mathcal{O}_{Y_b}^\dagger(\mathbf{p}, 0)}$ yields the spectral decomposition, which can be used as a model for fits to data:
    \begin{equation}\label{2ptsf}
        C_{Y_a \rightarrow Y_b}(\boldsymbol{p},t) = \sum_n \qty({(-1)^{t-1}})^n \sqrt{\frac{Z_a^{(n)}(\boldsymbol{p})}{2E_n(\boldsymbol{p})}} \Bigl(e^{-E_n(\boldsymbol{p})t} + e^{-E_n(\boldsymbol{p})(N_t-t)}\Bigr)  \sqrt{\frac{Z_b^{(n)}(\boldsymbol{p})}{2E_n(\boldsymbol{p})}}
    \end{equation}
    where $N_t$ is the temporal extent of the lattice, $Z^{(n)}$ and $E_n$ are, respectively, the overlap factor and the energy of the $n$th excited states, and the oscillating factor $(-1)^{t-1}$ arises due to the presence of particles with the opposite parity in the staggered discretization for the fermions. Most of the correlators appear in four different configurations depending on the combination of source and sink smearing ($d-d$, $1S-d$, $d-1S$ and $1S-1S$), and different momenta.
    Correlators related to collinear momenta $(p_x,0,0)$ come in two different orientations, parallel or perpendicular to the $D^{(*)}$ polarization. For each different momentum we perform a simultaneous fit of Eq.~\eqref{2ptsf} to the data corresponding to different smearings and momentum polarizations. 

    \subsubsection{Analysis of the systematics}

\begin{figure}
        \centering
        \includegraphics[width=0.7\linewidth]{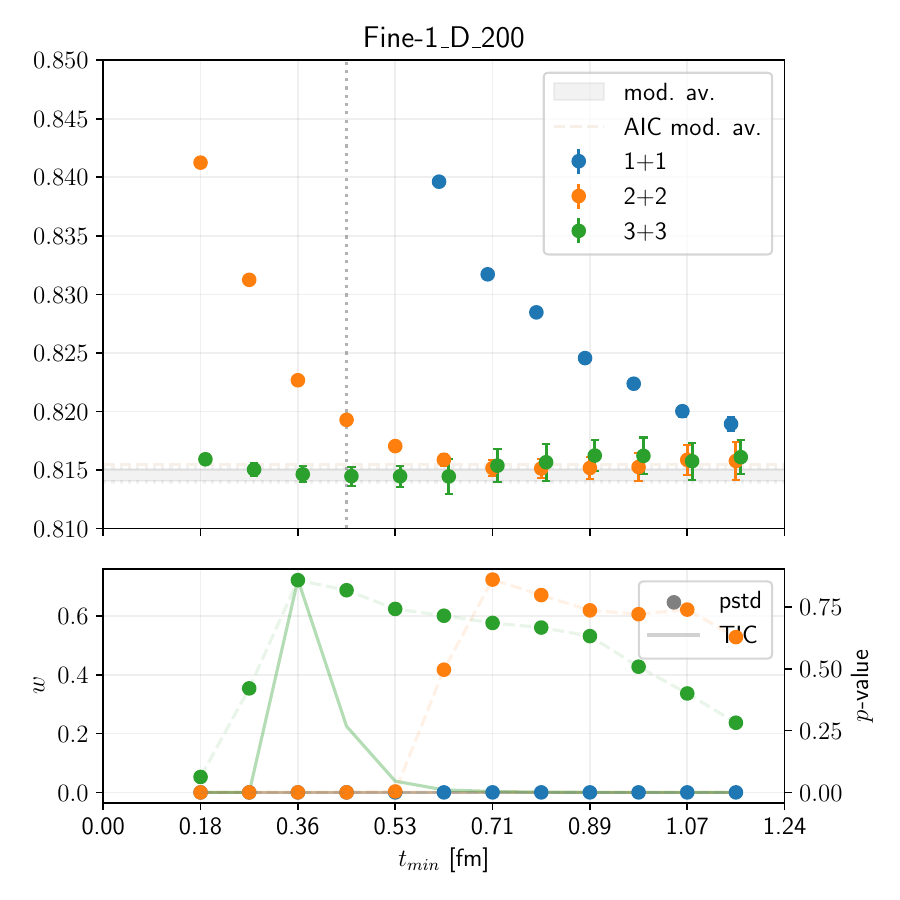}
        \caption{Fit results for $D$-meson correlator at momentum $(2,0,0)$ for $a=0.09$, $m_l/m_s=0.2$ for different values of the lower end of the fitting window $t_\text{min}$ and different number of excited states considered in the fitting model. Each dot corresponds to a value extracted from a different fit, each $x$-coordinate corresponds to a different value of $t_\text{min}$ each colour to a different $N_\text{states}$. \textit{Upper panel:} The $y$-axis correspond to the extracted value of the energy of the ground state. \textit{Lower panel:} each dot represents the $p$-value of the corresponding fit in the upper panel, the solid line to its TIC weight $w$ in the model average.}
        \label{2ptfit_stab}
    \end{figure}

    To handle autocorrelations, correlator data were binned and analyzed using jackknife resampling, applying the same fitting routines to each bin. A fully correlated fit was performed by minimizing a $\chi^2$ function with Gaussian priors, where the covariance matrix was rescaled using the non-binned dataset's correlation matrix. The shrinking procedure from Ref.~\cite{shrink}, also used in Ref.~\cite{alex}, was applied to the correlation matrix.
    The fitting hyper-parameters include the fitting window limits $[t_\text{min},t_\text{max}]$ and the number of states $N_\text{states}$ truncating the spectral sum in Eq.~\eqref{2ptsf}.\footnote{A ``$N_\text{states}+N_\text{states}$ fit" indicates equal numbers of physical and oscillating states in the model.} We use $N_\text{states}=3$ and aim to fix the fitting window in physical units: for $D$-meson fits, $[0.4,2.6] \text{ fm}$ is chosen, while for $D^{(*)}$ mesons, $t_\text{min}\sim 0.6$ fm with $t_\text{max}$ chosen to ensure a signal-to-noise ratio under 25\%.
    Systematic effects from hyper-parameter choices are analyzed by performing fits across various values of $t_\text{min}, t_\text{max},$ and $N_\text{states}$. The stability of extracted energies and overlap factors is assessed, focusing on various $t_\text{min}$ fits at fixed $t_\text{max}$. Stability regions are identified using the ground state's energy, $p$-value distributions~\cite{fsspval}, reduced $\chi^2$ values~\cite{chiexp}, and compatibility with the model-averaged results via the Takeuchi Information Criterion (TIC)~\cite{julien}.
    Fig.~\ref{2ptfit_stab} illustrates this stability analysis for the ground state $D$-meson energy $E_0$. The upper panel shows $E_0$ values for various $t_\text{min}, t_\text{max},$ and $N_\text{states}$. The lower panel plots $p$-values and TIC weights. Results confirm $E_0$ from a $3+3$ fit is stable across $t_\text{min}$ values. For lower $N_\text{states}$, $t_\text{min}$ must increase to achieve stability, as signaled by $p$-values.

\subsection{3-point function analysis}
\begin{figure}
        \centering
        \includegraphics[width=1.\linewidth]{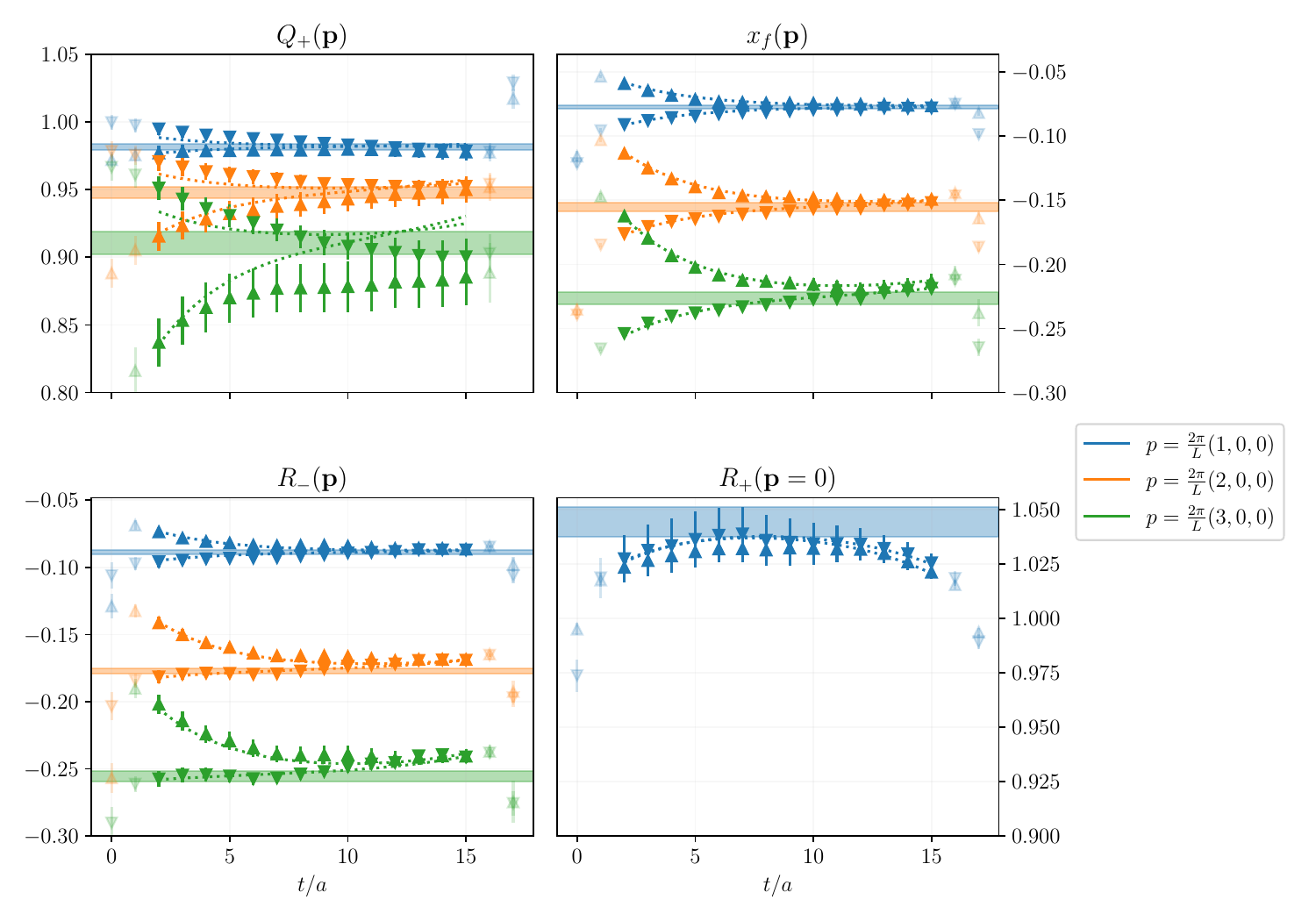}
        \caption{Results of the global fit of the relevant ratios for the $B\rightarrow D$ process for the same ensemble of the example case in Fig.\ref{2ptfit_stab}. The dashed lines indicate the fitted functional forms projected to every case, while the colored bands represent the value of $F_0$ with its error. We observe that the fit results are insensitive to the fitting window's choice. Each color represents a different momentum as in the legend, except for the panel depicting $R_+$, which is defined at 0 momentum.}
        \label{fig:ratiofitD}
    \end{figure}

\begin{figure}
        \centering
        \includegraphics[width=1.\linewidth]{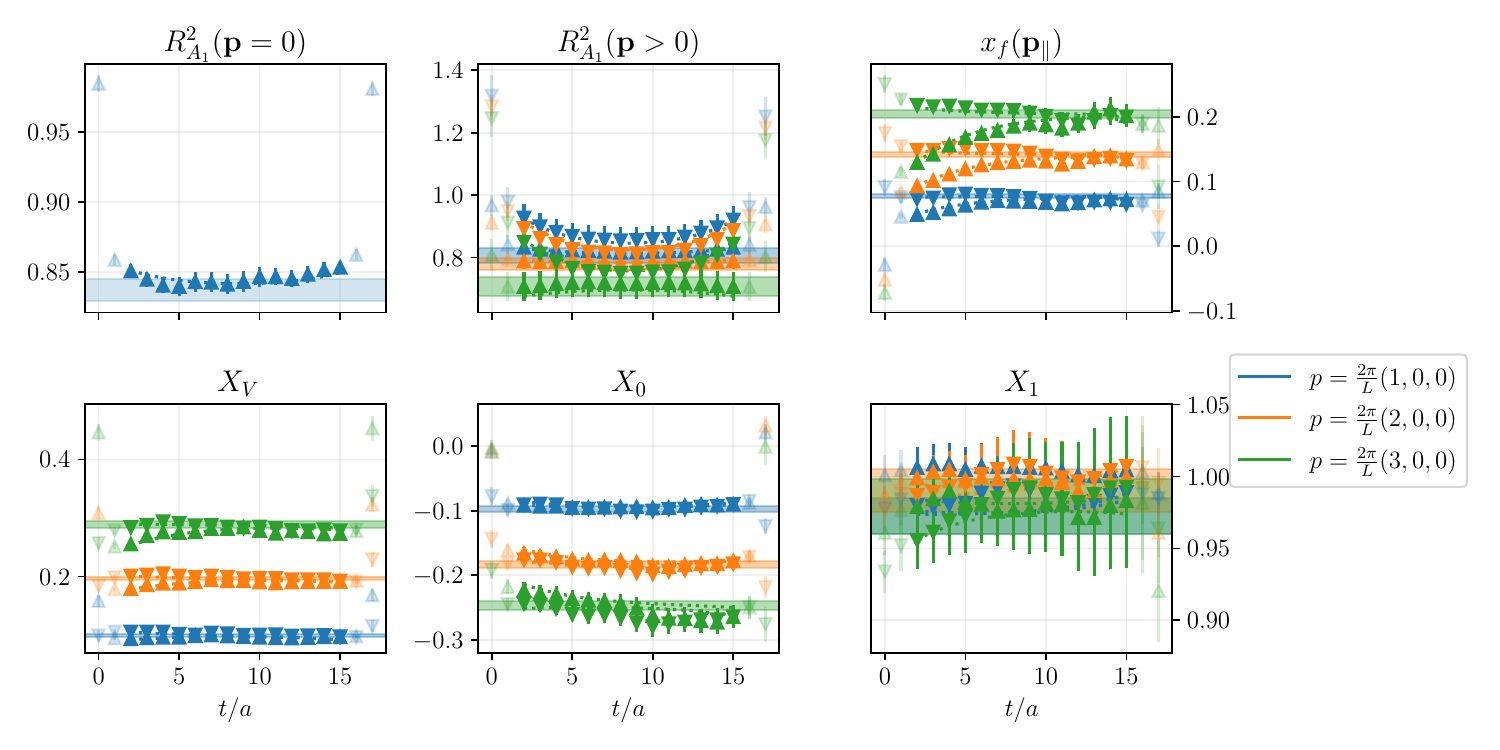}
        \caption{Same as in Fig.~\ref{fig:ratiofitD}, but for the $B\rightarrow D^*$ process. The legend also explains the colour code, except for the case of $R_{A_1}$, which is only defined at 0-momentum.}
        \label{fig:ratiofitDst}
    \end{figure}

    Using the same notation for Eq.~\eqref{2ptsf}, the 3-point correlators are constructed by inserting a current $J^\mu$ between the two meson interpolators as $\expval{\mathcal{O}_{Y_b}(0,T) J^{\mu}(p,t) \mathcal{O}^\dagger_{X_a}(-p,0)}$ (for a particular source-sink separation $T$ and time $t$). The spectral decomposition for the three-point function reads
    \begin{equation}\label{3pts}
        C_{X_a \rightarrow Y_b}^{J^\mu}(\boldsymbol{p}, t)=\sum_{n,m} s_n(t) s_m(T-t) \sqrt{Z_{Y_b, n}(\boldsymbol{p})} \frac{e^{-E_n(\boldsymbol{p}) t}}{2 E_n(\boldsymbol{p})}
            \left\langle Y_b, n, \boldsymbol{p}\left|J^\mu\right| X_a, m, \mathbf{0}\right\rangle \sqrt{Z_{X_a, m}(\mathbf{0})} \frac{e^{-M_m(T-t)}}{2 M_m},
    \end{equation}
    where $s_n(t)$ denotes the oscillating phase. Following the construction above, we build the different ratios $R(\boldsymbol{p},t,T)$ from Eq.~\eqref{ratioD} and ~\eqref{ratioDst} for 2 different source-sink separations $T$. As suggested in Ref.~\cite{smoothing}, we smooth the oscillating terms by computing the following quantity at each momentum
    \begin{equation}
        R(t,T) = \frac{1}{2}R(t,T) + \frac{1}{4}R(t,T+1)  + \frac{1}{2}R(t+1,T+1)
    \end{equation}
    We then perform a simultaneous fit to all the data corresponding to different ratios and different momenta using the following fitting model
    \begin{equation}\label{ratio}
        R(\boldsymbol{p},t,T) = F_0 + A(\boldsymbol{p}) e^{-\Delta E_{\text{source}}t} + B(\boldsymbol{p}) e^{-\Delta E_{\text{sink}}(T-t)}
    \end{equation}
    where $\Delta E$ is the difference between the energy of the first excited state and the ground state of the meson in the source and in the sink. This functional form comes from an expansion of Eq.~\eqref{3pts} to include the presence of excited states at the leading order. Using such a model, we can fit Eq.~\eqref{ratio} simultaneously to all our data (separately for the $B\rightarrow D$ and $B\rightarrow D^*$ case) with shared $\Delta E_{D^{(*)}}$ and $\Delta M_B$,\footnote{One can easily see that $\Delta E_{D^{(*)}}$ and $\Delta M_B$ are the only parameters of interest in the limit in which $p^2\ll M^2$ and $\Delta E_{D^{(*)}}\simeq \Delta E_{B}$, which we test to be the case for our data.}
    The results of the global fits fit for one ensemble are depicted in Fig.~\ref{fig:ratiofitD},~\ref{fig:ratiofitDst}.

    \subsection{Extraction of the form factor}
\begin{figure}
        \centering
        \includegraphics[width=1.\linewidth]{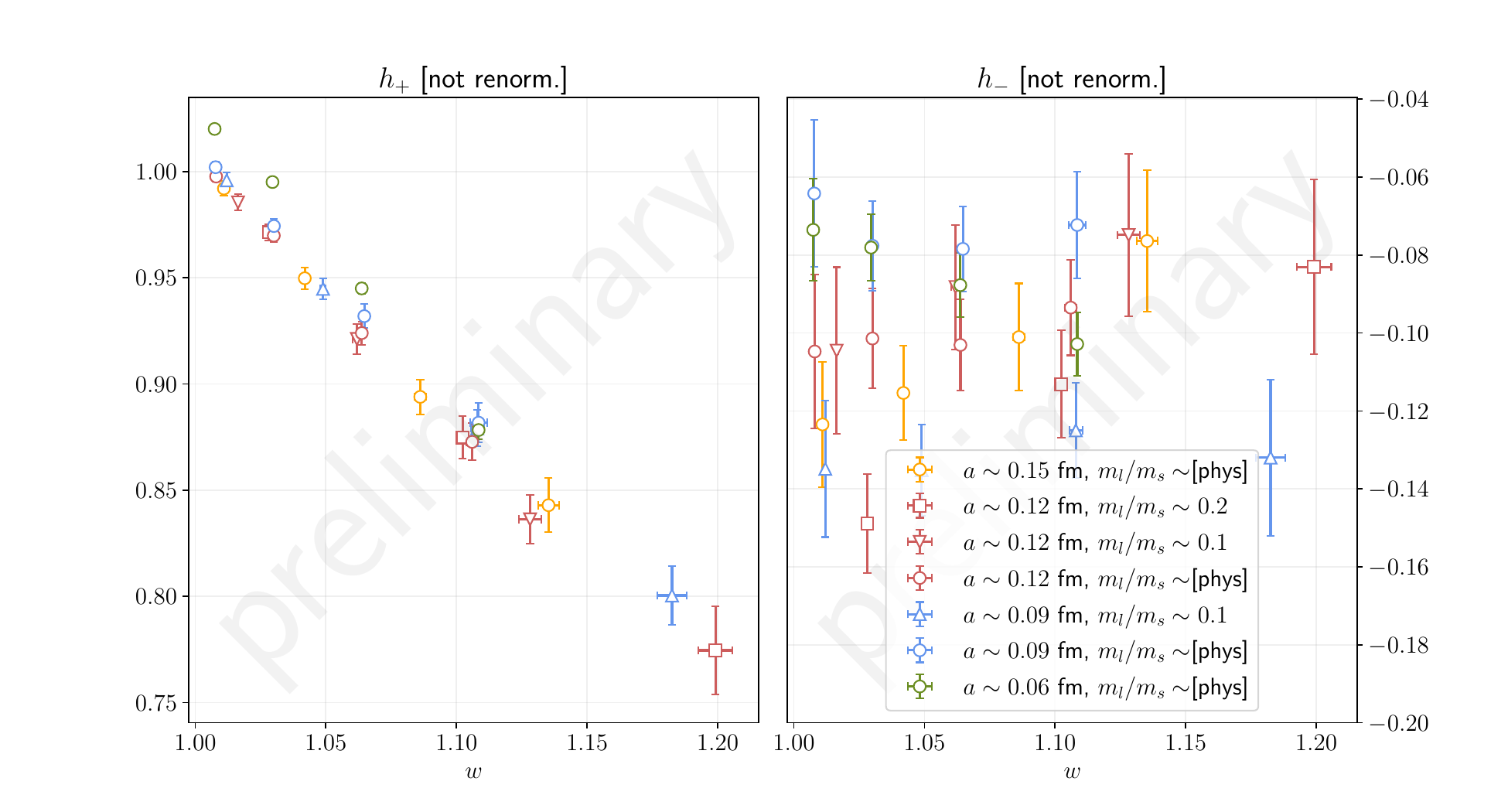}
        \caption{Bare, blinded form factors for the $B\rightarrow D$ process, defined in Eq.~\eqref{effD}.}
        \label{ffD}
    \end{figure}

\begin{figure}
        \centering
        \includegraphics[width=1.\linewidth]{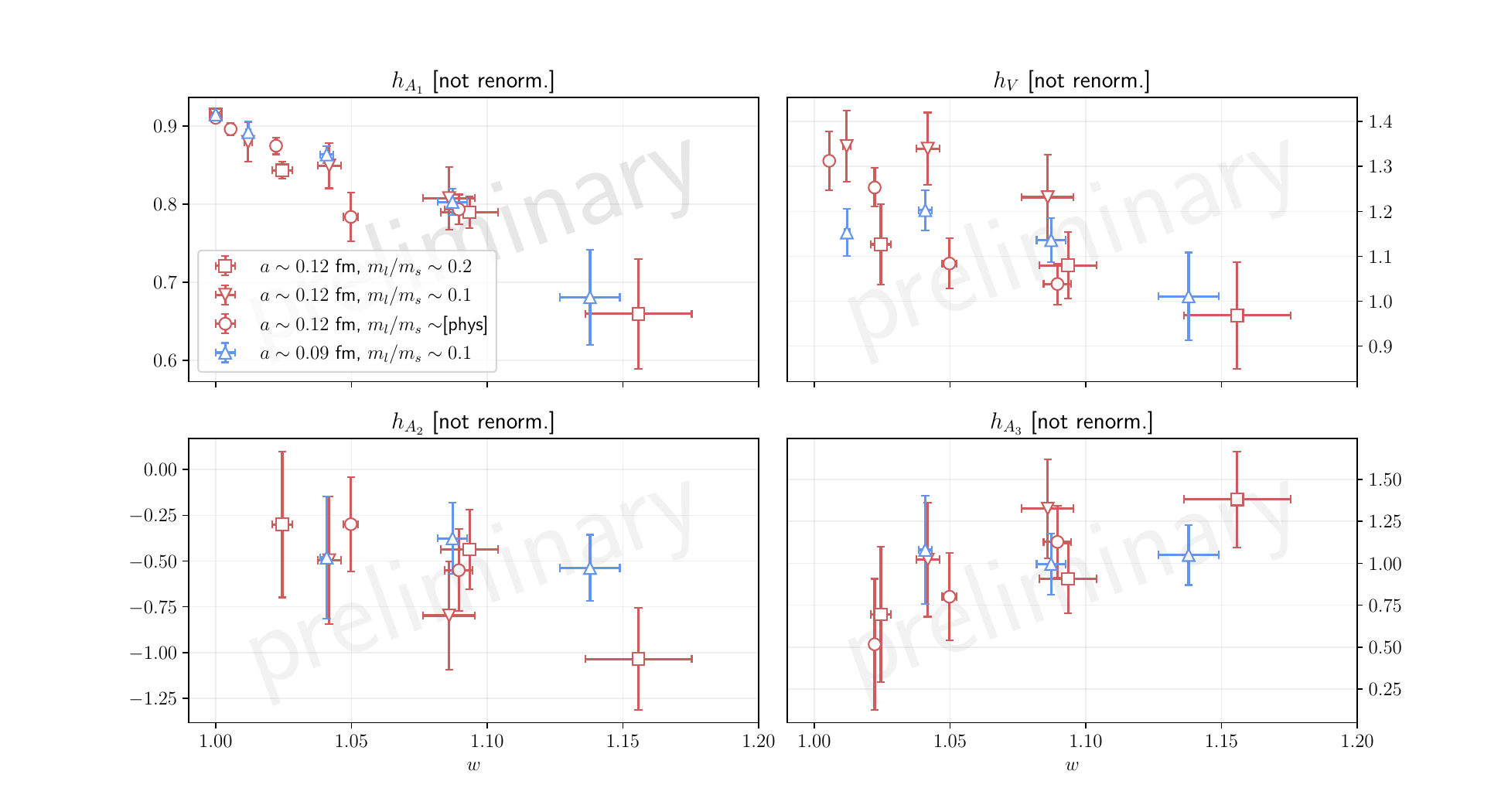}
        \caption{Bare, blinded form factors for the $B\rightarrow D^{*}$ process, for the ensemble indicated in the legend, defined in Eq.~\eqref{effDst}}
        \label{ffDst}
    \end{figure}

    We use the values of the ratio computed as in the previous subsection to build the form factors defined in Eq.~\eqref{effD} and ~\eqref{effDst}.
     In the form factor definitions, the barred ratios require renormalization. As the renormalization factors were unavailable at the time of this work, the final form factor values could not be constructed. However, since these factors are typically close to 1, plotting the non-renormalized form factor remains instructive.
     We plot the final preliminary results of this analysis in Fig.~\ref{ffD},~\ref{ffDst}.

\section{Conclusion}

We presented preliminary results from the FNAL-MILC collaboration's analysis of $B\to D^{(*)}\ell\nu$ decays, focusing on form factor extraction.. Using two- and three-point correlation data from $N_f=2+1+1$ MILC ensembles, we applied systematic uncertainty control and extracted the value of the ratios needed to compute the necessary form factors. The results show stable form factors and a good control over systematic effects. Future work includes continuum and physical mass extrapolations to finalize the analysis, advancing lattice QCD in semileptonic $B$ decays and precision flavor physics.

\newpage
\acknowledgments
This work was partly supported by the Grant DGA-FSE grant 2020-E21-17R Aragon Government and the European Union - NextGenerationEU Recovery and Resilience Program on “Astrofísica y Física de Altas Energías" CEFCA-CAPA-ITAINNOVA (P.B.); by the U.S. Department of Energy, Office of Science, under, No. DE-SC0010120 (S.G.), No. DE-SC0015655 (A.X.K., A.T.L.); 
by the National Science Foundation under Grants Nos. PHY20-13064 and PHY23-10571 (A.V.), and Grant No. 2139536 for Characteristic Science Applications for the Leadership Class Computing Facility (H.J.) by the Simons Foundation under their Simons Fellows in Theoretical Physics program (A.X.K.); 
by the Exascale Computing Project (17-SC-20-SC), a collaborative effort of the U.S. Department of Energy Office of Science and the National Nuclear Security Administration (H.J.); 
by the Universities Research Association Visiting Scholarship awards 20-S-12 and 21-S-05 (S.L.); by MICIU/AEI/10.13039/501100011033 and FEDER (EU) under Grant PID2022-140440NB-C21 (E.G.); 
by MCIN/AEI/10.13039/501100011033/FEDER, UE under Grants No. PID2019-106087GB-C21 and PID2022-140440NB-C21 (E.G.); by the Junta de Andalucía (Spain) under Grant No. FQM-101 (E.G.); 
by Consejeria de Universidad, Investigacion e Innovacion and Gobierno de Espana and EU – NextGenerationEU, under Grant AST22 8.4 (E.G.); by AEI (Spain) under Grant No. RYC2020-030244-I / AEI / 10.13039/501100011033 (A.V.); DE-SC0010120 (S.G.), DE-SC0011090 (W.J.), DE-SC0021006 (W.J.); by the U.S. National Science Foundation under Grants No. PHY17-19626 and PHY20-13064 (C.D., A.V.); and by U.K. Science and Technology Facilities. 

Computations for this work were carried out with resources provided by the USQCD Collaboration; by the ALCF and NERSC, which are funded by the U.S. Department of Energy; and by OLCF (Summit), NERSC(Cori, Edison), FNAL LQCD clusters, TACC. Fermilab is operated by Fermi Research Alliance, LLC under Contract No. DE-AC02-07CH11359 with the United States Department of Energy, Office of Science, Office of High Energy Physics.

\bibliographystyle{JHEP}
\bibliography{bibliography}

\providecommand{\href}[2]{#2}\begingroup\raggedright\begin{thebibliography}{10}

\bibitem{FLAG}
{\scshape Flavour Lattice Averaging Group (FLAG)} collaboration, \emph{{FLAG Review 2024}},  \href{https://arxiv.org/abs/2411.04268}{{\ttfamily 2411.04268}}.

\bibitem{bonn}
A.~Vaquero, \emph{{$B\to D^{(\ast)}\ell\nu$ semileptonic decays at non-zero recoil}}, \href{https://doi.org/10.22323/1.430.0250}{\emph{PoS} {\bfseries LATTICE2022} (2022) 250} [\href{https://arxiv.org/abs/2212.10217}{{\ttfamily 2212.10217}}].

\bibitem{alex}
{\scshape Fermilab Lattice and MILC} collaboration, \emph{{Semileptonic form factors for $B\rightarrow D^*\ell \nu $ at nonzero recoil from $2+1$-flavor lattice QCD: Fermilab Lattice~and~MILC~Collaborations}}, \href{https://doi.org/10.1140/epjc/s10052-022-10984-9}{\emph{Eur. Phys. J. C} {\bfseries 82} (2022) 1141} [\href{https://arxiv.org/abs/2105.14019}{{\ttfamily 2105.14019}}].

\bibitem{ensemble}
{\scshape Fermilab Lattice and MILC} collaboration, \emph{{Lattice QCD Ensembles with Four Flavors of Highly Improved Staggered Quarks}}, \href{https://doi.org/10.1103/PhysRevD.87.054505}{\emph{Phys. Rev. D} {\bfseries 87} (2013) 054505} [\href{https://arxiv.org/abs/1212.4768}{{\ttfamily 1212.4768}}].

\bibitem{milcD}
{\scshape Fermilab Lattice and MILC} collaboration, \emph{{B\textrightarrow{}D\ensuremath{\ell}\ensuremath{\nu} form factors at nonzero recoil and |V$_{cb}$| from 2+1-flavor lattice QCD}}, \href{https://doi.org/10.1103/PhysRevD.92.034506}{\emph{Phys. Rev. D} {\bfseries 92} (2015) 034506} [\href{https://arxiv.org/abs/1503.07237}{{\ttfamily 1503.07237}}].

\bibitem{smearing1}
J.L.~Richardson, \emph{{The Heavy Quark Potential and the Upsilon, J/psi Systems}}, \href{https://doi.org/10.1016/0370-2693(79)90753-6}{\emph{Phys. Lett. B} {\bfseries 82} (1979) 272}.

\bibitem{smearing2}
{\scshape Fermilab Lattice, MILC} collaboration, \emph{{B- and D-meson decay constants from three-flavor lattice QCD}}, \href{https://doi.org/10.1103/PhysRevD.85.114506}{\emph{Phys. Rev. D} {\bfseries 85} (2012) 114506} [\href{https://arxiv.org/abs/1112.3051}{{\ttfamily 1112.3051}}].

\bibitem{shrink}
O.~Ledoit and M.~Wolf, \emph{Improved estimation of the covariance matrix of stock returns with an application to portfolio selection}, \href{https://doi.org/10.1016/s0927-5398(03)00007-0}{\emph{Journal of Empirical Finance} {\bfseries 10} (2003) 603–621}.

\bibitem{fsspval}
D.~Toussaint and W.~Freeman, \emph{{Sample size effects in multivariate fitting of correlated data}},  \href{https://arxiv.org/abs/0808.2211}{{\ttfamily 0808.2211}}.

\bibitem{chiexp}
M.~Bruno and R.~Sommer, \emph{{On fits to correlated and auto-correlated data}}, \href{https://doi.org/10.1016/j.cpc.2022.108643}{\emph{Comput. Phys. Commun.} {\bfseries 285} (2023) 108643} [\href{https://arxiv.org/abs/2209.14188}{{\ttfamily 2209.14188}}].

\bibitem{julien}
J.~Frison, \emph{{Towards fully bayesian analyses in Lattice QCD}},  \href{https://arxiv.org/abs/2302.06550}{{\ttfamily 2302.06550}}.

\bibitem{smoothing}
{\scshape Fermilab Lattice and MILC Collaborations} collaboration, \emph{$\overline{B}\ensuremath{\rightarrow}{D}^{*}\ensuremath{\ell}\overline{\ensuremath{\nu}}$ form factor at zero recoil from three-flavor lattice qcd: A model independent determination of $|{V}_{cb}|$}, \href{https://doi.org/10.1103/PhysRevD.79.014506}{\emph{Phys. Rev. D} {\bfseries 79} (2009) 014506}.

\end{thebibliography}\endgroup

\end{document}